\documentclass[11 pt, preprint,preprintnumbers,amsmath,amssymb, prd]{revtex4}

\usepackage{latexsym}
\usepackage{amssymb}
\usepackage{epsfig,amsmath,graphics}
\usepackage{epstopdf}

\newcommand{\GeV}{\text{GeV}}

\newcommand{\angstrom}{\buildrel _{\circ} \over {\mathrm{A}}}

\newcommand{\MGUT}{M_{\text{GUT}}}
\newcommand{\MPl}{M_{\text{Pl}}}

\def\r{\right)}
\def\l{\left(}

\begin{document}


\title{Axion Dark Matter Detection with Cold Molecules}

\author{Peter W. Graham}
\affiliation{Stanford Institute for Theoretical Physics, Department of Physics, Stanford University, Stanford, CA 94305}

\author{Surjeet Rajendran}
\affiliation{Department of Physics and Astronomy, The Johns Hopkins University, 3400 N. Charles Street, Baltimore, MD 21218, USA}

\date{\today}

\begin{abstract}
Current  techniques cannot detect axion dark matter over much of its parameter space, particularly in the theoretically well-motivated region where the axion decay constant $f_a$ lies near the GUT and Planck scales.  We suggest a novel experimental method to search for QCD axion dark matter in this region. The axion field oscillates at a frequency equal to its mass when it is a component of dark matter. These oscillations induce  time varying CP-odd nuclear moments, such as electric dipole and Schiff moments. The coupling between internal atomic fields and these nuclear moments gives rise to time varying shifts to atomic energy levels. These effects can be enhanced by using elements with large Schiff moments such as the light  Actinides, and states with large spontaneous parity violation, such as molecules in a background electric field. The energy level shift in such a molecule can be $\sim 10^{-24}$ eV or larger.  While challenging, this energy shift may be observable in a molecular clock configuration with technology presently under development. The detectability of this energy shift is enhanced by the fact that it is a time varying shift whose oscillation frequency is set by fundamental physics and is therefore independent of the details of the experiment.  This signal is most easily observed in the sub-MHz range, allowing detection when $f_a$ is $\gtrsim 10^{16}$ GeV, and possibly as low as $10^{15}$ GeV.  A discovery in such an experiment would not only reveal the nature of dark matter and confirm the axion as the solution to the strong CP problem, it would also provide a glimpse of physics at the highest energy scales, far beyond what can be directly probed in the laboratory.
\end{abstract}

\maketitle
\tableofcontents

\section{Introduction}
\label{Sec:Introduction}

The nature of dark matter is one of the great mysteries of our time. Extensive observational efforts have yielded direct empirical evidence for the particle nature of dark matter  \cite{Clowe:2006eq}. It is natural to expect a dark matter particle to have non-gravitational interactions. The discovery of such interactions will not only unveil the properties of dark matter but also offer a unique window into physics beyond the standard model. These interactions may also allow for detection of the dark matter in laboratory experiments. For example, following the pioneering work of \cite{Goodman:1984dc}, there are now several experiments that are aimed towards detecting  weakly interacting massive particles (WIMPs).  One other dark matter candidate is generally considered to be as well-motivated as the WIMP: the axion.  However, while there are many experiments that can detect WIMPs, there are  few that can search for axion dark matter.  In fact, there is not even a proposed method to detect axion dark matter over most of its parameter space.  In this paper we propose a new technique to search for axion dark matter in a previously inaccessible region of parameter space. 

We argue that the dark matter axion can affect atomic and molecular energy levels  (see section \ref{Sec: overview}). These energy differences can be measured very precisely through tools such as atomic and molecular interferometry.  The past decade has witnessed rapid developments in the field of atomic interferometry. Interferometers with the ability to measure phase shifts as small as $\sim \frac{10^{-4}}{\sqrt{\text{Hz}}}$  \cite{clocks, Metcalf}  are routinely constructed in laboratories. Furthermore, significant improvements in the sensitivities of these detectors are expected in the near future, owing to possible advances in atom cooling technologies, the development of techniques to beat the standard quantum limit in these interferometers (for example, through the use of squeezed states)  \cite{squeezedstate, Tuchman_PRA} and the increased ability to manipulate cold atom states \cite{Phillips2002:JPhysB, HolgerLMT, McGuirk}. Molecular interferometers have yet to achieve the same level of sensitivity as atom interferometers. However, techniques to significantly improve the sensitivity of such molecular interferometers seem possible in the near future \cite{ColdMolecules}. The impressive sensitivity of these devices and the many possible near term advances in their sensitivity makes atom and molecular interferometry a particularly attractive option to pursue in the hunt for new physics. 

In this article, we examine the use of atom and molecular interferometers in the detection of axion dark matter. This paper is organized as follows. In section \ref{Sec: axion dark matter}, we briefly overview the phenomenology of axion dark matter, including a discussion of current axion detection strategies and their limitations. Section \ref{Sec: overview} gives the basic idea of our proposal. Following this discussion, section \ref{Sec:Strategy} gives the details of the experimental strategy. Finally, in section \ref{Sec:Conclusions}, we conclude. 

\section{Axion Dark Matter}
\label{Sec: axion dark matter}
The axion was initially introduced as a solution to the strong CP problem \cite{Peccei:1977hh, Peccei:1977ur,Weinberg:1977ma, Wilczek:1977pj, Kim:1979if, Shifman:1979if, Dine:1981rt, Zhitnitsky:1980he}.  The Strong CP problem is essentially the fact that the term
\begin{equation}
\mathcal{L} \supset  \frac{g^2_s}{32 \pi^2} \theta_\text{QCD} \, \text{tr } G \tilde{G}
\label{Eqn:theta term}
\end{equation}
where $G$ is the field strength of QCD and $g_s$ is the QCD gauge coupling, is allowed by all symmetries in the Lagrangian of the Standard Model (SM) and yet experimental observations limit $\theta_\text{QCD} \lesssim 6 \times 10^{-10}$.  Due to the term in Eqn.~\eqref{Eqn:theta term} the neutron acquires a dipole moment $d_n$ of \cite{Pospelov:1999ha}
\begin{equation}
\label{eqn:theta dipole moment}
d_n = 1.2 \times 10^{-16} \, \theta_{\text{QCD}} \, \text{e} \cdot \text{cm}
\end{equation}
The current bound on $\theta_{\text{QCD}}$ arises from experimental bounds on the dipole moment of the neutron  \cite{Pospelov:1999ha}.  In the SM there is no reason for $\theta_\text{QCD}$ to be so small, and thus the Strong CP problem is taken as motivation for physics beyond the SM.

\begin{figure}
\begin{center}
\includegraphics[width = 5.5 in]{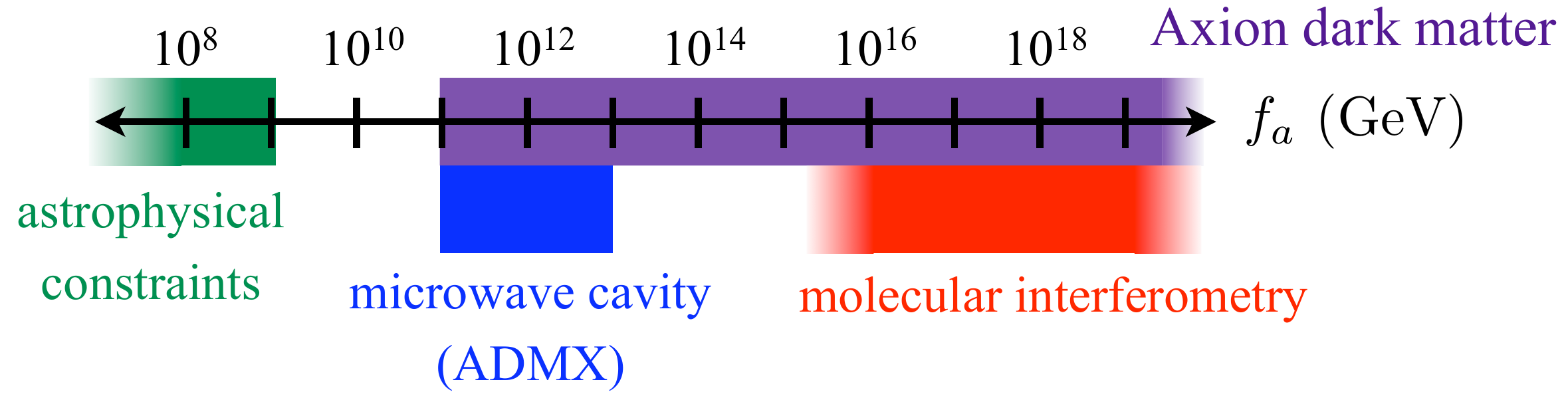}
\caption{ \label{Fig:parameterspace} (Color Online) The parameter space of the axion in $f_a$ (GeV).  Values of $f_a < 10^9$ GeV are ruled out by astrophysical constraints (green).  Values of $f_a \gtrsim 10^{11}$ GeV allow the axion to be the dark matter (purple).  The (blue) region labelled ``microwave cavity" shows the region of parameter space that is potentially observable with microwave cavity experiments, e.g.~ADMX.  The (red) region labelled ``molecular interferometry" shows the range of $f_a$ which is potentially observable with the experiments proposed here.  The lower limit on this region may in fact be lower than shown, depending on technological advances.}
\end{center}
\end{figure}

The Peccei-Quinn mechanism  \cite{Peccei:1977hh, Peccei:1977ur} elegantly solves this problem by turning the parameter $\theta_\text{QCD}$ into a dynamical field, the axion $a(x)$.  Over cosmological time the axion redshifts towards the minimum of its potential, dynamically setting the $\theta$ parameter to zero.  The axion has a potential which is conventionally written as
\begin{equation}
\mathcal{L} \supset \frac{g^2_s}{32 \pi^2} \frac{a}{f_a} \text{tr } G \tilde{G} + m^2_a a^2
\label{Eqn:axiongluon}
\end{equation}
where $f_a$ is the axion decay constant, and the axion's mass is given by
\begin{equation}
\label{eqn:axion mass}
m_a \sim \frac{\Lambda^{2}_{\text{QCD}}}{f_a}
\end{equation}
where $\Lambda_\text{QCD} \sim 200$ MeV is the QCD confinement scale.  The axion arises as the pseudo Goldstone boson of the breaking of a global $U(1)_{PQ}$ symmetry which has a mixed anomaly with QCD \cite{Peccei:1977hh, Peccei:1977ur,Weinberg:1977ma, Wilczek:1977pj}.  This symmetry is broken at the scale $f_a$ which then suppresses all the couplings of the axion to the SM and determines the phenomenology of the axion \cite{Peccei:1977hh, Peccei:1977ur,Weinberg:1977ma, Wilczek:1977pj, Kim:1979if, Shifman:1979if, Dine:1981rt, Zhitnitsky:1980he}.  Axion parameter space is shown in Figure \ref{Fig:parameterspace}.

Theoretically, the scale $f_a$ is expected to lie near the fundamental scales $\MGUT \l \sim 10^{16} \, \GeV \r  \, - \, \MPl \l \sim 10^{19} \, \GeV \r$ of particle physics. This is a natural expectation since many theories of physics beyond the standard model involve the breaking of gauge symmetries (like the grand unified group) and  space-time symmetries (through compactification) at these scales. Additionally, the well known incompatibility of global symmetries with quantum gravity also suggests that these fundamental scales are a likely place for the scale ($f_a$) of the breaking of the $U(1)_{PQ}$ symmetry.  Another of the many motivations for the axion is the ease with which it can arise from string theory, or any theory with nontrivial extra dimensions \cite{Svrcek:2006yi}.  In this case, $f_a$ is naturally near these high scales, at the GUT or string scales.

Stringent astrophysical and other laboratory bounds rule out the axion parameter space in the region  $f_a \lessapprox 10^{9}$ GeV  \cite{Dicus:1978fp, Vysotsky:1978dc, Raffelt:2006cw}, as shown in Figure \ref{Fig:parameterspace}. In the astrophysically allowed region ($f_a \gtrapprox 10^{9}$ GeV), it was quickly realized that the axion could have cosmological relevance \cite{Preskill:1982cy, Abbott:1982af, Dine:1982ah}. The initial misalignment angle of the axion field from its minimum contributes to the energy density of the universe. In particular, it was pointed out in \cite{Preskill:1982cy, Abbott:1982af, Dine:1982ah} that the axion can easily overclose the universe when $f_a \gtrapprox 10^{12}$ GeV. This result was initially regarded as evidence against $f_a$ lying in the theoretically favored window  $\MGUT  -  \MPl$, since the axion can exist in this window only if its initial misalignment angle was tuned to be small.   However, in fact it is basically the product of the initial misalignment angle and $\frac{f_a}{\MPl}$ which must be small.  Thus there is no particular reason to believe that the initial misalignment angle must be large while $\frac{f_a}{\MPl}$ is small.  Further, it was realized that a piece of the universe, easily large enough to be our observable universe would naturally emerge with such a small initial misalignment angle if the universe had undergone a period of inflation.  Since inflation naturally generates a universe which is significantly larger than our observable universe, the natural expectation is to get many different large patches which have many different values of the initial misalignment angle.  Though such a region may seem ``small", that is in fact an ill-defined statement and the right question, the probability of being in such a region, is measure-dependent.  If for example, there were some anthropic constraints on the dark matter density then it seems quite likely to find ourselves  in a region of the universe which has the observed dark matter density, even if $f_a$ is far above $10^{12}$ GeV \cite{Linde:1987bx, Freivogel:2008qc}.  Even without anthropic constraints the probability of being in such a region is ill-defined. In this context, it is natural to expect an axion with $f_a \gtrapprox 10^{12} \, \GeV$ to be a significant component of dark matter \cite{Linde:1987bx}. 

The window $\MGUT \lessapprox f_a \lessapprox \MPl$ is thus a theoretically well motivated region of the axion parameter space where the axion can easily be a significant component of dark matter. While there are a variety of experiments probing the axion window in the region between $10^{10} \text{ GeV} \lessapprox f_a \lessapprox 10^{12} \text{ GeV}$  \cite{Sikivie:1985yu,  Asztalos:2009yp}, there are no known methods to probe $f_a$ between $\MGUT \lessapprox f_a \lessapprox \MPl$ in a laboratory (for an interesting astrophysical probe, see \cite{Arvanitaki:2009fg}). 

Current axion dark matter searches  \cite{Sikivie:1985yu,  Asztalos:2009yp} for  $10^{10} \text{ GeV} \lessapprox f_a \lessapprox 10^{12} \text{ GeV}$  aim to detect the conversion of axions into photons in an appropriately designed microwave cavity.  This axion-photon conversion happens in a background magnetic field through the coupling
\begin{equation}
\mathcal{L} \supset g_{a\gamma} \frac{a}{f_a} F \tilde{F} = g_{a\gamma} \frac{a}{f_a} \vec{E} \cdot \vec{B}
\label{Eqn:axionphotons}
\end{equation}
where $F$ is the electromagnetic field strength, $g_{a \gamma}$ depends upon the details of the axion model \cite{Kim:1979if, Shifman:1979if, Dine:1981rt, Zhitnitsky:1980he}, and $\vec{E}$ and $\vec{B}$ are the electric and magnetic fields.
 There are significant challenges in extending these detection techniques to higher values of $f_a$. Since the coupling in Eqn.~\eqref{Eqn:axionphotons} is suppressed by $f_a$, the axion-photon conversion rate scales as $\sim \frac{1}{f^2_a}$, resulting in a drastic loss in  signal for higher values of $f_a$. Additionally, these experiments also rely on a resonance between the axion mass $m_a \sim 500 \text{ m}^{-1} \l \frac{\MGUT}{f_a}\r$ and the length of the cavity in order to enhance the signal to measurable levels. Since the mass of the axion is unknown, the experiment has to scan through different cavity lengths in order to search for and resonate with the axion. The construction and alteration of such cavities is increasingly difficult for high $f_a$ axions as the physical size $\sim 500 \text{ m}  \l \frac{f_a}{\MGUT}\r$ of the cavity needed to achieve resonance becomes prohibitively large. Owing to these difficulties, it is desirable to develop alternate methods to search for axions with $f_a \gg 10^{12} \, \GeV$.

 \section{Overview of the Detection Method}
 \label{Sec: overview}
 
 In this article we propose a method for detecting axions with high $f_a$, in the range $f_a \gtrsim 10^{16} \text{ GeV}$.  This could even be used to detect axions with $f_a$ above the Planck scale $\sim 10^{19}$ GeV, though such axions may be theoretically difficult to construct \cite{Banks:2003sx}.  The basic idea is to detect the small shifts in atomic energy levels caused, as we shall discuss, by the background axion dark matter.  We will use the technology of atomic clocks since this is the best way to measure atomic energy splittings.


All previous axion searches, in particular the microwave cavity experiments, rely on the coupling in Eqn.~\eqref{Eqn:axionphotons}.  This is a derivative coupling - hence, its physical effects are suppressed by the ratio of the size of the probe and the axion wavelength (often squared).  For atomic and molecular probes, this is a severe suppression since the probe sizes are $\sim 1$ angstrom while the axion wavelengths are $\sim 500 \text{ m } \l \frac{f_a}{\MGUT}\r$.

Our idea is to search for effects of the coupling of the axion to the gluon, Eqn.~\eqref{Eqn:axiongluon}, instead of the electromagnetic coupling.  This interaction is not a pure derivative owing to the QCD anomaly \cite{'tHooft:1976up, 'tHooft:1976fv}, which is why it can generate a nucleon electric dipole moment as in Eqn.~\eqref{eqn:theta dipole moment}
\begin{equation}
\label{eqn: axion dipole moment}
d_n = 1.2 \times 10^{-16} \, \frac{a}{f_a} \, \text{e} \cdot \text{cm}
\end{equation}
where $\theta_\text{QCD}$ has been replaced by the axion $\theta_{a} = \frac{a}{f_a}$.  Thus the background axion dark matter field generates an electric dipole moment (EDM) in every nucleon.

To calculate this EDM we must know the value of the axion dark matter field locally.  The ``vacuum misalignment" production mechanism is the dominant axion dark matter production mechanism \cite{Preskill:1982cy, Abbott:1982af, Dine:1982ah}.  It is based on the fact that the axion field started at an almost constant value everywhere in the observable universe after inflation.  The axion field is constant initially because inflation exponentially suppresses spatial gradients in the axion field.  Thus the axion begins as a classically coherent scalar field after inflation.  Of course, the dominant potential for the axion is given by its mass term in Eqn.~\eqref{Eqn:axiongluon}.  It is easy to solve the equation of motion with this quadratic potential and hence the axion dark matter field has a value of
\begin{equation}
\label{eqn: oscillating axion}
a(t) = a_0 \cos \left( m_a t \right)
\end{equation}
where $a_0$ is the constant amplitude of the axion.  The axion acts as a coherently oscillating classical scalar field because it started at a constant value everywhere in the observable universe after inflation \cite{Preskill:1982cy, Abbott:1982af, Dine:1982ah}.  Note that this means that the axion field starts with very small spatial gradients, thus a very low velocity, allowing it to be a good cold dark matter candidate even though it has a very low mass.

The classical axion field continues to oscillate as in Eqn.~\eqref{eqn: oscillating axion} because all the interaction terms in its Lagrangian are irrelevant operators suppressed by at least one power of the very large scale $f_a$, see for example Eqns.~\eqref{Eqn:axiongluon} and \eqref{Eqn:axionphotons}.  Thus the axion's interactions with itself and all other particles in the universe are extremely suppressed and all of its scattering cross sections are very small.  The axion field then never thermalizes with the rest of the matter in the universe since it essentially does not interact with it.  This is important because if it thermalized it could not possibly be a cold dark matter candidate.  Without interactions, it thus remains a coherently oscillating classical scalar field as in Eqn.~\eqref{eqn: oscillating axion}.  Note that this is completely different from a thermally produced gas of axion particles at some temperature which would have no coherence.  While there is a component of thermally produced axions (from collisions of Standard Model particles), it is significantly smaller than this non-thermal component.  Thus this coherence naturally occurs over all of axion dark matter parameter space $f_a \gtrsim 10^{12} \, \GeV$, and indeed is responsible for the axion being a good cold dark matter candidate.

The only relevant effect on the axion field from the rest of the matter in the universe is during structure formation from the large-scale gravitational effects of the forming clumps of matter.  Although these large-scale effects certainly cannot cause the axion field to thermalize into a gas of particles, they do virialize the axion field, causing it to acquire the virial velocity of the object it is in, in our galaxy that is $v \sim 10^{-3}$.  Thus the classical axion field in Eqn.~\eqref{eqn: oscillating axion} picks up small spatial gradients
suppressed by a factor of $10^3$.
Of course then the coherence length is not infinite (as it would be if Eqn.~\eqref{eqn: oscillating axion} were exactly correct) but finite of order the do Broglie wavelength $\sim \frac{1}{m_a \, v}$ (set by the spatial gradients).  Equivalently, the coherence time is limited by the spread in kinetic energy of the axion $\sim m_a v^2 \sim 10^{-6} m_a$, and is thus about $10^6$ times the period of the axion field.  See Section \ref{Sec:Strategy} for more details.  Consequently, in an experiment that probes the dark matter distribution over length scales smaller than $\sim \frac{1}{m_a \, v} \sim 500 \text{ km } \l \frac{f_a}{\MGUT}\r$, the dark matter axion behaves like a coherently oscillating scalar field. It is also easy to see that given the very light mass of the axion, $m_a \sim 10^{-19}  \, \GeV \l\frac{\MGUT}{f_a}\r$, the dark matter density in our galaxy requires so many axions within each de Broglie wavelength that there must be a macroscopic occupation number of each axion mode.
Under the assumption that the axion is the dark matter, we can calculate the amplitude $a_0$ locally. Equating the energy density in the axion field with the local dark matter density $\rho_\text{DM}$
\begin{equation}
m_a^2 \, a^2 \sim \rho_\text{DM} \approx 0.2 \frac{\text{GeV}}{\text{cm}^3}
\end{equation}
gives a value for the axion amplitude of
\begin{equation}
\theta_{a} = \frac{a_0}{f_a} \sim \frac{\sqrt{\rho_\text{DM}}}{\Lambda^2_{QCD}} \sim 3 \times 10^{-19}
\label{Eqn:targettheta}
\end{equation}
Interestingly the induced value of $\theta_a$ is independent of $f_a$.

Thus, axion dark matter generates an electric dipole moment for every nucleon immersed in it which oscillates rapidly in time from Eqns.~\eqref{eqn: axion dipole moment}, \eqref{eqn: oscillating axion}, and \eqref{Eqn:targettheta} as
\begin{equation}
\label{eqn: oscillating dipole moment}
d_n \approx 4 \times 10^{-35}  \cos \left( m_a t \right) \, \text{e} \cdot \text{cm}
\end{equation}
This oscillation has a frequency given by the axion mass $m_a \approx \, 1 \, \text{kHz} \,  \l \frac{\MPl}{f_a} \r \,  \approx \, 1 \, \text{MHz} \, \l \frac{\MGUT}{f_a} \r$.  We propose a search for this oscillating EDM.

Naively, this appears to be a daunting task.  The current experimental bound on a {\it static} nucleon EDM is about $\sim$ 9 orders of magnitude larger than this value.  There are however  big differences between the methods used to search for static and rapidly oscillating EDMs.  In fact, the static nucleon EDM experiments take place on long timescales and would thus be completely insensitive to the oscillating EDM caused by the axion.  Interestingly, despite the fact that the axion arises from high energy physics, the frequency band of oscillation, kHz - MHz, is experimentally accessible.

There are essentially two reasons searches for an oscillating EDM can be more sensitive than searches for a static EDM.  The oscillation of the signal is at a frequency set by fundamental physics independent of the experiment.  Searching for such a resonance with a sharp peak in frequency space greatly reduces backgrounds.  Further, the oscillation allows the use of internal atomic electric fields, which are much larger than the fields that can be applied in the laboratory.  

Experimentally, the only way to observe a static EDM is to apply an external electric field $E_{\text{ext}}$ to an atom or neutron and look for energy shifts of the form $\vec{E}_{\text{ext}} \cdot \vec{d}_n$. The sensitivity of these experiments is therefore limited by the maximum electric field  $E_{\text{ext}}$ that can be safely created in the lab. 
However with an oscillating EDM, we can exploit the coupling to the internal electric fields of the atom $\vec{E}_{\text{int}}$ which gives rise to time dependent energy shifts of the atomic levels. These internal fields are several orders of magnitude larger than the largest fields we can construct in the lab. For example, even in a Hydrogen atom these fields can be as large as  $E_{\text{int}} \sim \frac{e}{\angstrom^2} \sim 10^{10} \frac{\text{V}}{\text{cm}}$. Naively, this electric field couples to the induced nuclear EDM giving rise to a time dependent energy shift. The exact value of this shift depends upon the atomic species used, but these are typically order (see Table \ref{Tab: actinides})
\begin{equation}
\delta E \sim  E_{\text{int}} \, d_n \sim 10^{-24} \text{ eV} - 10^{-25} \text{ eV} 
\label{Eqn:EnergyShift}
\end{equation}
The detection of this tiny shift would be extremely difficult if it was static since it would require extraordinary control over backgrounds. However, since the above energy shift oscillates at a relatively high frequency $\sim$ kHz - MHz, it should be easier to detect than a static energy shift (see section \ref{Sec:Backgrounds}). 


Measurements of energy shifts of this magnitude, while a factor of  $\mathcal{O}\l10^2 - 10^{4} \r$ beyond the reach of current atomic and molecular clocks/interferometers, can conceivably be achieved in the near future. We discuss the experimental strategy required to measure such an energy shift in section \ref{Sec:Strategy}. For the rest of this section, we will focus on sharpening the naive estimates of the energy shift made in equation \eqref{Eqn:EnergyShift}. 

The naive estimate of the energy shift $\delta E$ in \eqref{Eqn:EnergyShift} was made under the implicit assumption that the expectation value $\langle \Psi | \vec{E}_{\text{int}} | \Psi \rangle$ of the average internal electric field $\vec{E}_{\text{int}}$ in an experimentally accessible, meta-stable (for time scales $\sim 1$ s) state $|\Psi \rangle$ of the atom or molecule was non-zero. This assumption requires refinement. In particular, there are two conditions that are necessary in order for $\langle \Psi | \vec{E}_{\text{int}} | \Psi \rangle$ to be non-zero. 

First,  since the electric field $\vec{E}_{\text{int}}$ is a vector, it is parity (P) odd. Hence,  a state $| \Psi \rangle$ can have a non-zero expectation value for it only if the state breaks parity. In the absence of parity violation, this vector does not have a direction to point to and hence its average value  $\langle \Psi | \vec{E}_{\text{int}} | \Psi \rangle$ is zero. Now, since electromagnetism conserves parity, the energy eigenstates of atoms and molecules are also parity eigenstates. This implies that in such an energy eigenstate, the state $| \Psi \rangle$ is either parity odd or even, and hence the expectation value $\langle \Psi | \vec{E}_{\text{int}} | \Psi \rangle$ of a parity odd  operator $\vec{E}_{\text{int}}$ is zero. 

However, laboratory states $|\Psi_L \rangle$ of atoms and molecules that break parity explicitly can be created by application of an external electric field $\vec{E}_{\text{ext}}$. The external field $\vec{E}_{\text{ext}}$ then explicitly breaks parity and the energy eigenstates of the atom or molecule exposed to this electric field will also explicitly break parity. In such a state $| \Psi_L \rangle$, the expectation value $\langle \Psi_L | \vec{E}_{\text{int}} | \Psi_L \rangle$ is proportional to the polarization induced by the external field $\vec{E}_{\text{ext}}$. A molecule can be essentially completely polarized by the application of moderate electric fields $\sim 10 - 100 \, \frac{\text{kV}}{\text{cm}}$ \cite{ColdMolecules, Hinds:1976ph}. The ease of molecular polarizability is easy to understand. For example, in a polar, linear, diatomic molecule, the average molecular dipole moment is $\sim$ e-angstrom. This molecular dipole moment couples to the external field  $E_{\text{ext}} \sim 100 \, \frac{\text{kV}}{\text{cm}}$ giving rise to energy shifts $\sim 10^{-3}$ eV. The rotational levels in a molecule are also at a similar energy and hence these rotational modes get maximally mixed by the applied field, leading to a complete polarization of the molecule. The molecular axis of the state $| \Psi_L \rangle$ is then aligned along the direction of the external field $\vec{E}_{\text{ext}}$ giving rise to an  expectation value $\langle \Psi_L | \vec{E}_{\text{int}} | \Psi_L \rangle$ that is roughly $\sim  \vec{E}_{\text{int}}$. It is important to note that even though an external electric field  $\vec{E}_{\text{ext}}$  is applied to create the state $|\Psi_L \rangle$,  $\langle \Psi_L | \vec{E}_{\text{int}} | \Psi_L \rangle$ is still roughly equal to $ \vec{E}_{\text{int}} \gg \vec{E}_{\text{ext}}$. The estimate \eqref{Eqn:EnergyShift} for the energy shift caused by the dark matter axion would then still hold for such a state $|\Psi_L\rangle$. The independence of $\langle \Psi_L | \vec{E}_{\text{int}} | \Psi_L \rangle$ on the applied electric field $\vec{E}_{\text{ext}}$ is of course only true in the range $E_{\text{ext}} \gtrapprox 10 - 100 \, \frac{\text{kV}}{\text{cm}}$ where the polarization of the molecule is saturated. 

Similarly, polarized atomic states where the average value of the internal field does not vanish can be created by application of external electric fields. However, since the atom (unlike a molecule) does not have an intrinsic, permanent dipole moment, the interactions of the external field with the atom are all second order. Furthermore, the rotational levels of the atom are split by $\sim$ eV instead of the  $\sim 10^{-3}$ eV splittings of rotational molecular states. Consequently, significant polarization of atomic states requires large external fields $\sim 10^{10} \, \frac{\text{V}}{\text{cm}}$. These are difficult to achieve under laboratory conditions.  In the absence of such fields, the estimate \eqref{Eqn:EnergyShift} for the energy shift will be suppressed by the polarization fraction of the atom in the applied field. Since observing the energy shift in Eqn.~\eqref{Eqn:EnergyShift} is already challenging, we probably cannot afford significant suppressions of \eqref{Eqn:EnergyShift} due to the small polarizability of atoms. Consequently, for the rest of this paper, we will focus our attention on the energy shift  \eqref{Eqn:EnergyShift} caused by the dark matter axion on molecules in the presence of moderate external electric fields. We note however that  the size of the external field required to polarize an atom can be ameliorated by a careful choice of atomic species which has low lying rotational levels $\ll 1$ eV (for example, in Radium \cite{Flambaum:1999zz}). We do not pursue this option in this paper. 

We note that the naive estimate \eqref{Eqn:EnergyShift} could have also been applied to a nuclear energy shift caused by the dark matter axion field {\it i.e.} the internal electrical fields in the nucleus $\sim e \,  \Lambda^{2}_{QCD}$ could have interacted with the neutron dipole moment to give rise to a time varying energy shift of the nucleus. Since the nuclear electric fields are $\sim$ 10 orders of magnitude larger than atomic electric fields, such time dependent energy shifts would seem to be an attractive venue to pursue. However, as  in the case of electromagnetism, the nuclear ground state needs to break P in order for this energy shift to have a non-zero value. Additionally, since the axion is  CP odd, it can give rise to a non-zero energy shift only if the state also breaks T.  Since the strong interactions conserve P, T, the only other source of P, T violation in the nucleus is from the weak interactions. While the weak interactions violate P maximally, the violations of T are CKM suppressed and are very small. Consequently, any such possible nuclear energy shift is severely suppressed by the smallness of T violation in nuclear physics. This is unlike the case of the polarized molecule where owing to the application of the external electric field, the molecule did not have internal axes under which it had well defined parity properties. Consequently, the polarized molecular state also does not have well defined properties under P,T thereby enabling the CP violating axion field to induce an energy shift in the molecule. But, one cannot significantly polarize the nucleus by the application of external electric fields owing to its small size and the large ($\sim$ keV - MeV) binding energies of nuclear rotational levels. Hence, nuclear energy shifts caused by the axion are negligibly small. 

With the application of moderate external electric fields on a molecule, we can create states $| \Psi_L \rangle$ that have the right P, T properties to allow $\langle \Psi_L | \vec{E}_{\text{int}} | \Psi_L \rangle$ to be non-zero. However, there is an additional hurdle that needs to be overcome. We require  $\langle \Psi_L | \vec{E}_{\text{int}} | \Psi_L \rangle$ evaluated at the position of the nucleus to be non-zero since it is that electric field which couples to the neutron dipole moment. But, the electric field at the location of the nucleus in a dominantly electrostatic system like a molecule is suppressed by virtue of Schiff's theorem \cite{Schiff}. This theorem states that in a system at equilibrium held together by electrostatic forces, the average electric field at the location of each charged, point particle must vanish. Physically, this arises because for a system in equilibrium, the average force on each particle must vanish and in the case of a system held together by electrostatic forces, the electric field is the only source of such forces. In the absence of such an electric field, the energy shift due to the neutron dipole moment is zero. However, since the nucleus has a finite size, the Schiff theorem can be evaded through interactions between the electric field and higher nuclear moments. 

The most significant of these interactions, particularly for the case of high $Z$ atoms, is the so-called Schiff moment \cite{Flambaum:1984fb}. The Schiff moment can be used to compute an effective dipole moment $d_S$ that couples to the electric field $\vec{E}_{\text{int}}$  \cite{Schiff, Flambaum:1984fb}, leading to an energy shift $\sim d_S \, E_{\text{int}}$. This effective dipole moment $d_S$ is roughly $\sim  \l 10^{-9} \r Z^3$ \cite{Flambaum:1984fb} of the initial neutron dipole moment. In order to minimize this suppression, it is therefore desirable to use molecules consisting of atoms with very large $Z$. In conventional searches for the static neutron dipole moment, molecules of Mercury and Thallium are often used. For these systems, $d_S \sim 10^{-3} \, d_n$ \cite{Dmitriev:2004gk}. If molecules of Mercury or Thallium are used to search for the dark matter axion, then the energy shift \eqref{Eqn:EnergyShift} would have to be suppressed by an additional factor of $\sim 10^{-3}$ in order to account for the Schiff suppression. Since the energy shift \eqref{Eqn:EnergyShift} is already experimentally challenging, it is desirable to search for other nuclei where the Schiff moments are not as badly suppressed. 

It has been recently realized that there are nuclei where specific properties of the nuclear structure allows for significantly larger Schiff moments than those of Mercury or Thallium. In particular, the actinides, whose nuclear structure have octupole deformations and possess low lying, opposite parity levels, have significantly enhanced Schiff moments \cite{Auerbach:1996zd, Spevak:1996tu}. For example, the Schiff moment of Radium 225 is $\sim$ 400 times larger than that of Mercury, while the enhancement in Protactinium 229 is almost a factor of $10^{4}$ \cite{Spevak:1996tu}.  For the case of Radium 225, the effective dipole moment $d_S \sim 0.4$ times the neutron dipole moment $d_n$, while for Protactinium 229, $d_S \sim 10$ times larger than $d_n$. Similar enhancements are also expected for Francium.  Consequently, if molecules containing these elements are used to run the interferometer, the Schiff suppression essentially disappears. In this case,  \eqref{Eqn:EnergyShift} essentially is the correct energy shift caused by the dark matter axion. 

\begin{table}
\begin{center}
\begin{math}
\begin{array}{|c|c|c|c|}
\hline
\text{Element} & \text{Suppression Factor} & \text{Energy Shift} & \text{Half-life} \\
\hline
^{225}\text{Ra} & 0.2 & \sim 2 \times 10^{-25} \text{ eV} & 15 \text{ d} \\
\hline
^{239}\text{Pu} & 0.3 & \sim 3 \times 10^{-25} \text{ eV} & 2.4 \times 10^4 \text{ yr} \\
\hline
^{223}\text{Fr} & 0.4 & \sim 4 \times 10^{-25} \text{ eV} & 22 \text{ min} \\
\hline
^{225}\text{Ac} & 0.6 & \sim 6 \times 10^{-25} \text{ eV} & 10 \text{ d} \\
\hline
^{229}\text{Pa} & 9 & \sim 9 \times 10^{-24} \text{ eV} & 1.4 \text{ d} \\
\hline
\end{array}
\end{math}
\caption[Elements]{\label{Tab: actinides} A few examples of isotopes with largest Schiff moments.  These Schiff moments were calculated in \cite{Spevak:1996tu}. For each one the associated suppression factor is the factor by which the naive estimate for the energy shift due to the dipole moment in Eqn.\eqref{Eqn:EnergyShift} must be multiplied.  Note that in the case of $^{229}$Pa this is in fact an enhancement.  The corresponding energy shift is also given, as is the half-life of the isotope.  The energy shift is necessarily approximate, as described in the text, since the actual energy shift is also determined by the choice of molecule.}
\end{center}
\end{table}

The energy shifts induced by the dark matter axion for various actinide nuclei are tabulated in Table \ref{Tab: actinides}.  These are some of the best known elements for this search.  Others elements that are known to have similar enhancement factors can be found in the literature (see e.g.~\cite{Auerbach:1996zd, Spevak:1996tu}) and of course it may still be found that other elements work as well or better than these. These nuclei are all unstable, but their lifetimes are long enough to allow interferometry over time scales $\sim 1$ s. The use of such highly radioactive materials in an experiment is of course non trivial. However, we note that Radium 225 is currently being used in an atom interferometer setup that is aimed at searching for the static dipole moment of the neutron \cite{Scielzo:2006xc}. The use of Radium 225 in \cite{Scielzo:2006xc} is precisely motivated by the large enhancement in its Schiff moment relative to that of Mercury, which also helps the search of static nuclear dipole moments. This experiment aims to trap as many as $\sim 10^6$ laser cooled Radium 225 atoms per second in its quest to extend the reach of EDM experiments by a factor of 100 - 1000. Although the chemical characteristics of a Radium molecule are different from that of a Radium atom, the difficulties associated with dealing with a radioactive nucleus are similar to both a molecular and an atomic interferometer. Consequently, the progress achieved by \cite{Scielzo:2006xc} in dealing with such radioactive nuclei should be useful in the construction of a molecular interferometer that uses Radium. The difficulties of using a highly radioactive source like Radium 225 can be ameliorated through the use of more stable nuclear isotopes that nevertheless possess comparable Schiff moment enhancements.  For example, it has been noted \cite{Engel:1999np} that the long lived isotope Plutonium 239 possesses soft octupole deformations in its nuclear structure resulting in the enhancement of Schiff moments by a factor of 300 relative to Mercury.  Molecules using Plutonium can also therefore be useful in an experiment to search for the dark matter axion. 

For the above reasons, in this paper, we now focus on the possibility of using a molecular interferometer to detect the dark matter axion. We will assume that the interferometer can be run with molecules containing an actinide nucleus ({\it e.g.} Radium, Plutonium etc.). Of course, the actinide nucleus needs to carry a non-zero nuclear spin, so that the dark matter axion can create a dipole moment for the nucleus which can then interact with the internal fields of the molecule. With the application of an external electric field $\sim 100 \, \frac{\text{kV}}{\text{cm}}$, the molecule can be maximally polarized to a state $| \Psi_L  \rangle$ in which the molecular axis is oriented along the direction of the external electric field. The internal electric fields $\sim \frac{e}{\angstrom^2} \sim 10^{10} \frac{\text{V}}{\text{cm}}$, that are way bigger than the applied external field,   interact with the Schiff moment of the actinide nucleus in the molecule giving rise to energy shifts  $\sim 10^{-24}$ eV, as given by  \eqref{Eqn:EnergyShift}. The exact value of this energy shift depends upon the choice of actinide nucleus (see Table \ref{Tab: actinides}). After incorporating various $\mathcal{O} \l 1 \r$ factors, this energy shift is $\sim 10^{-25}$ eV for molecules containing Radium 225 and Plutonium 239.  In a molecule with Protactinium, the energy shift is larger and is $\sim 9 \times 10^{-24}$ eV. This energy shift oscillates with a frequency given by the axion mass $m_a \sim$ kHz - MHz. In the following sections, we discuss experimental strategies to measure these time dependent energy shifts. For concreteness, we will quote numbers for molecules containing Radium 225 or Plutonium 239 and hence our target energy sensitivity will be $\sim 10^{-25}$ eV.

\section{Experimental Strategy} 
\label{Sec:Strategy}
In this section, we discuss an experimental strategy to measure the time varying energy shift $\sim 10^{-25}$ eV in a molecular interferometer. This section is organized as follows: we first discuss some general features of the strategy which includes details like the ability to convert this energy shift into a phase shift and the technique to allow the induced phase shift to add coherently over several axion oscillation periods. Following this prelude, we summarize the concrete details of the proposed experiment in sub section \ref{Sec:Setup}  and the required sensitivity of the interferometer in sub section \ref{Sec:Sensitivity}.  We also briefly discuss some experimental backgrounds in sub section \ref{Sec:Backgrounds}, although the discussion is by no means exhaustive.  In sub section \ref{Sec:Technology}, we discuss the technological advances needed to achieve the sensitivity goals of sub section \ref{Sec:Sensitivity} and finally, in sub section \ref{Sec:Reach} we discuss the region of the dark matter axion parameter space probed by this technique. 

We begin by illustrating a method to convert the energy shift \eqref{Eqn:EnergyShift} into a phase shift.  For concreteness, we consider a linear, polar, diatomic molecule although our analysis will hold for a general polar molecule. Under the application of an external electric field, this molecule is polarized with its permanent, molecular dipole moment aligned along the direction of the external field. In such a molecule,  the energy shift \eqref{Eqn:EnergyShift} caused due to the interaction between the nuclear dipole moment induced by the dark matter axion and the internal fields of the molecule depends upon the relative orientation between the nuclear dipole  moment and the polarized molecular axis. The nuclear dipole moment is aligned along the direction of the nuclear spin. Hence, a phase shift can be observed in a molecule if the molecule is placed in a linear superposition with two states $| \Psi_L \rangle_a$ and  $| \Psi_L \rangle_o$: $| \Psi_L \rangle_a$ where the nuclear spin is aligned along the polarized molecular axis and $| \Psi_L \rangle_o$ in which the nuclear spin is anti-aligned (see figure \ref{Fig:setup}). The energy shifts in the two states will be of opposite sign, leading to a relative phase shift between them, which can then be measured. 

Since the energy shifts created by the dark matter axion are tiny, it is necessary to interrogate over several periods of the axion oscillations in order to accumulate a sufficiently large phase in the interferometer. In order to do so, we need to rotate the nuclear spins with respect to the molecular axis at the frequency of the axion rotation $m_a \sim 1 \text{kHz - MHz}$ so that the phase difference accumulated as a result of the energy oscillation coherently adds for as long as possible. This can be achieved by placing the molecules in a suitable magnetic field $\vec{B}_{\text{ext}}$, resulting in the precession of the nuclear spin. Precession frequencies $\sim 1 \text{ MHz} \, \l\frac{B_{\text{ext}}}{0.1 \text{ T}}\r$ can be achieved with laboratory size fields. Of course, since we do not know the mass of the axion, the experiment has to scan through the frequency band of interest in its search for the axion. This can be  achieved in this experiment by changing the strength of the external field. Unlike the case of current cavity searches for axion dark matter, where the physical size of the cavity needs to be altered in order to look for a different axion frequency, in this experiment the frequency scanning can be done relatively easily. For example, if an electromagnet is used to provide the requisite magnetic field, then this magnetic field can be changed by altering the current through the magnet. We emphasize that the precession of the nuclear spin through the application of an external magnetic field is just one way to alter the relative orientation between the nuclear spin and the molecular axis. Other methods could also be used to achieve this goal, for example, through the application of suitable RF waves that can induce spin-flips. 

It is worth noting that the ease with which this experiment can scan through different axion frequencies is fundamentally because the experiment measures the non-derivative coupling \eqref{Eqn:axiongluon} of the axion with nuclei. Hence, the physical size of the apparatus does not play a role in determining the strength of the axion interactions. This is unlike the case of cavity searches for dark matter, where the derivative coupling \eqref{Eqn:axionphotons} of the axion is used. In order for a derivative coupling to be unsuppressed, the physical size of the apparatus must be comparable to the axion wavelength. In such an experiment, scanning the axion band without loss in signal requires alteration of the cavity size of the experiment. The ability of the proposed molecular interferometer to easily scan through various axion frequencies is particularly useful in searches for axions in the band  $\MGUT \lessapprox f_a \lessapprox \MPl$ where the limitations of current cavity based axion searches are evident. 

Ideally, we would like to interrogate the molecule for the maximum time $\tau_{\text{max}}$ possible, since the accrued phase shift is directly proportional it. Theoretically, the upper limit on this time is set by the coherence time of the axion dark matter field. Beyond this coherence time, the axion field behaves like noise and consequently the phase shift will not add  linearly.  The coherence time scale of the experiment is set by the time taken by the experiment to traverse the spatial gradients of the classical dark matter axion field. For the dark matter axion, these spatial gradients are $\sim \frac{1}{m_a v}$ \cite{Preskill:1982cy, Abbott:1982af, Dine:1982ah} where $v \sim 10^{-3}$ is the virial velocity of the dark matter in the galaxy. Since the relative velocity between the Earth and the galactic dark matter is also the virial velocity $v$, the maximum coherence time for the experiment is 
\begin{equation}
\tau_{\text{max}} =  \frac{1}{m_a \, v^2} \sim \, 1 \, \text{s} \, \l \frac{f_a}{\MGUT}\r
\label{Eqn:MaxTime}
\end{equation}
Of course, practical considerations can make the interrogation times smaller than the maximum time \eqref{Eqn:MaxTime} allowed by the coherence of the axion field. For example, if the molecular interferometer is operated by allowing the molecules to be in free fall during the course of the interferometry, then the physical size of the experiment sets a limitation on the interrogation time of the experiment. The molecule falls by roughly $\sim \frac{1}{2} \, g \, T^2 \,  \sim 5 \, \text{m} \, \l\frac{T}{\text{1 s}}\r^2$ in an interrogation time $T$ in such an interferometer. Interferometers of this physical size have been proposed and constructed in other precision searches  \cite{Dimopoulos:2006nk, Dimopoulos:2007cj, Dimopoulos:2008hx, Dimopoulos:2008sv} and hence achieving interrogation times $T \sim 1$ s seems feasible in a free fall molecular interferometer. However, interrogation times much longer than $\sim 1$ s are difficult to achieve in a free fall interferometer as
 the physical size of the interferometer rapidly becomes large. As we will see in sub section \ref{Sec:Sensitivity}, it is desirable to achieve interrogation times  as large as possible ( $T >> 1$ s), particularly while searching for high $f_a$ axions. Such interrogation times could be realized if the molecules are held in a suitable trap. In this case, the trap life time would set an upper bound on the interrogation time of the experiment.

It is evident from the above discussion that the experiment relies crucially on the ability to control and manipulate nuclear spins. If the nuclear spins were all aligned in random directions with respect to the molecular axis, the effect of the axion induced energy shift will average out over the sample. Additionally, we also need to control the spin in order to make it precess at the axion frequency $m_a$. Both the initial alignment of the spins and the nuclear spin precession can be achieved by the application of moderate laboratory magnetic fields $\sim 1$ T. However, this control scheme is effective only if the molecules are sufficiently cold. The energy splitting induced by the interaction of a $\sim 1$ T magnetic field with the nuclear spin is $\sim 10^{-7}$ eV $\sim 1$ milli kelvin. Consequently, the molecules must be cooled to at least milli kelvin temperatures. We note that the experiment also requires control over the molecular axis and this is to be achieved with the application of an external electric field $\sim 100 \, \frac{\text{kV}}{\text{cm}}$. Since the interaction energy between this external electric field and the permanent molecular dipole moment $e-$angstrom is $\sim 10^{-3}$ eV $\gg 1$ milli kelvin, when the molecules are cooled to these milli kelvin temperatures, they will be automatically aligned with the external electric field. We also note that the size of the molecular cloud will expand due to its temperature. With a milli kelvin cloud and an interrogation time $T \sim 1$ s, the cloud size is $\sim 30$ cm $\l\frac{T}{\text{1 s}} \r$. This cloud size is well within laboratory scales. However, if longer interrogation times are to be achieved (see sub section \ref{Sec:Sensitivity}), then cloud will have to be appropriately cooled. 

With these considerations in mind, we now discuss the experimental setup required to search for the dark matter axion. 

\subsection{The Setup} 
\label{Sec:Setup}
\begin{figure}
\begin{center}
\includegraphics[width = 5.5 in]{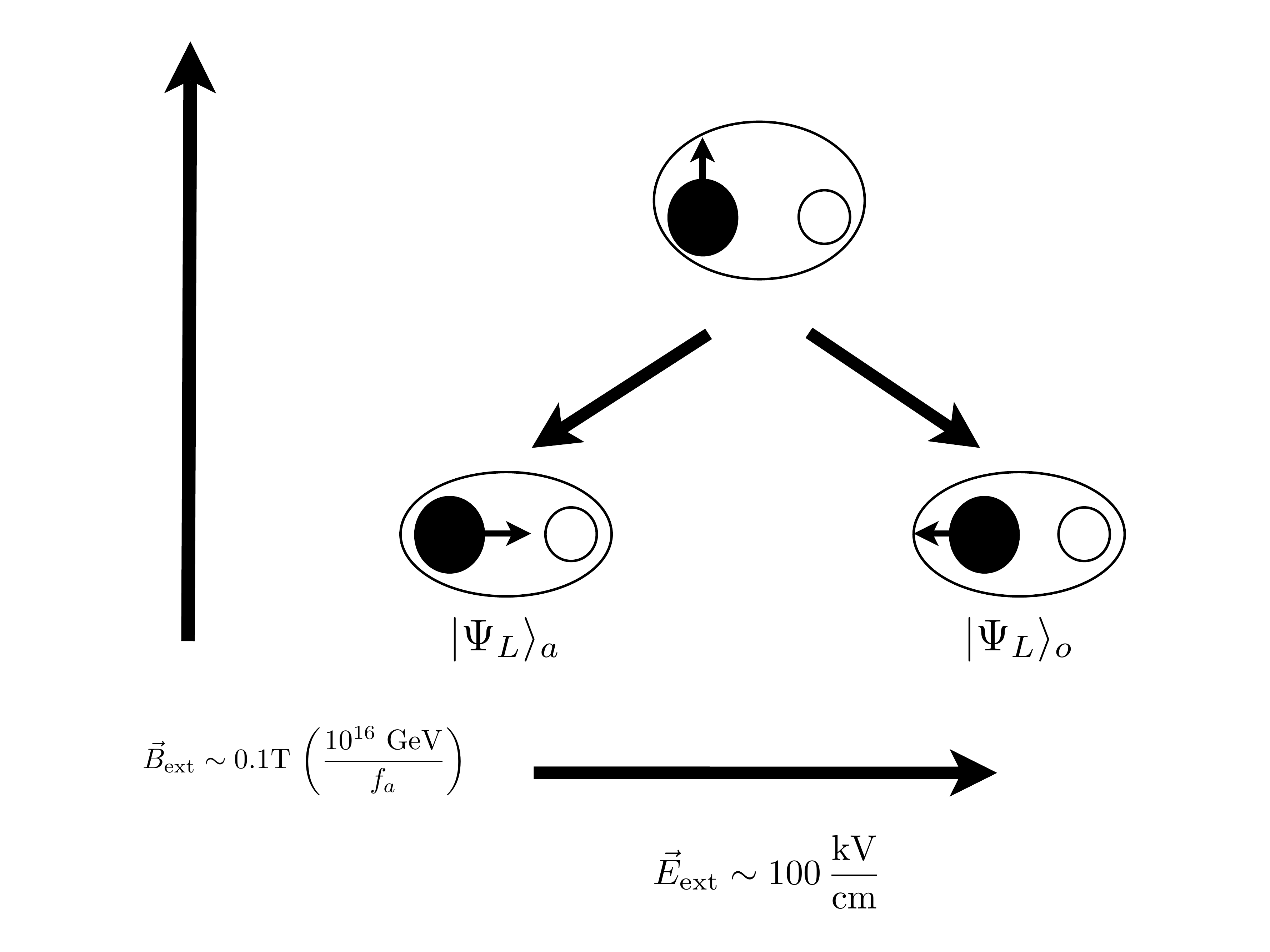}
\caption{ \label{Fig:setup} The molecules are polarized by an external electric field $\vec{E}_{\text{ext}} \sim 100 \, \frac{\text{kV}}{\text{cm}}$. They are then placed in a linear superposition of the two states $| \Psi_L \rangle_a$ and $| \Psi_L \rangle_o$, where the nuclear spin is either aligned or anti-aligned with the molecular axis respectively, leading to a phase difference between them in the presence of the axion induced nuclear dipole moment $d_n$.  The external magnetic field $\vec{B}_{\text{ext}} \sim 0.1 \text{ T} \, \l \frac{f_a}{\MGUT}\r$ causes the spins to precess, so that the phase difference can be coherently accrued over several axion oscillations.  The frequency can be scanned by dialing this magnetic field $\vec{B}_{\text{ext}}$ until it is resonant with the axion frequency.}
\end{center}
\end{figure}

We begin with an ensemble of polar molecules of some actinide element ({\it e.g.} Radium monohydride, RaH, with actinide radium possessing a non-zero spin), cooled to at least milli kelvin temperatures. It is necessary to align the molecular spin along some convenient direction through the application of an external magnetic field.  As discussed above, for a sample at milli kelvin temperatures, this requires  magnetic fields $\sim 1$ T. The strength of the required field is correspondingly smaller if lower molecular temperatures can be achieved. For the purposes of this discussion, we will take the direction of the magnetic field to lie along the vertical (see figure \ref{Fig:setup}). An external electric field $\sim 100 \, \frac{\text{kV}}{\text{cm}}$ must then be applied in a direction perpendicular to the magnetic field, resulting in a polarization of the molecular axis along that direction. In this proposal, this would be an electric field that lies along the horizontal (see figure \ref{Fig:setup}).

The molecules are now prepared in the appropriate initial state $| \Psi_L \rangle$ where the nuclear dipole moment can interact in a controlled way with the internal molecular electric fields.  After this, the molecules must be brought to the interferometer region  (see figure \ref{Fig:setup}). The molecular axis must still remain fixed and polarized, so the electric field in this interferometer region must also be   $\sim 100 \, \frac{\text{kV}}{\text{cm}}$ . For ease of description, we will assume that this electric field is in the same direction as the field that was used to prepare the molecular states. In this experiment, this is therefore an electric field oriented in the horizontal direction. The interferometer region also requires a magnetic field that must be perpendicular to the direction of the electric field, {\it i.e.} along the vertical  (see figure \ref{Fig:setup}). The axion frequency $m_a$ sets the magnitude of this magnetic field. This field needs to be $\sim 0.1 \text{ T} \, \l\frac{\MGUT}{f_a}\r$ in order for the precession frequency of the nuclear spin to match the axion frequency $m_a \sim 1 \text{ MHz} \, \l \frac{\MGUT}{f_a}\r$. Of course, the experiment can search for a different value of the axion mass by changing the strength of this external magnetic field. 

Once the molecules are ferried to this region, the actinide nucleus in the molecule must be placed in a linear superposition of two states $| \Psi_L \rangle_a$ and $| \Psi_L \rangle_o$, where the nuclear spin is either aligned or anti-aligned with the molecular axis respectively (see figure \ref{Fig:setup}).  This superposition can be created using traditional NMR techniques. The energy shift caused by the dark matter axion in the two states is of opposite sign and hence a phase difference will accrue between them. Furthermore, since the nuclear spins are now perpendicular to the applied magnetic field, they will precess with a frequency  set by the external magnetic field. When the precession frequency matches the axion frequency, a phase shift will be continually accrued over several axion oscillations. After interrogation for a time $T$, the phase shift in the experiment (using the energy shift $\delta E$ from \eqref{Eqn:EnergyShift}) is

\begin{equation}
\delta \phi = \delta E  \, T \sim 10^{-10} \, \l  \frac{T}{\text{1 s}}\r \, \l \frac{\delta E}{10^{-25} \text{ eV}} \r
\label{Eqn:PhaseShift}
\end{equation}

This relative phase between the two spin states  $| \Psi_L \rangle_a$ and $| \Psi_L \rangle_o$ can then be measured. 

\subsection{Sensitivity} 
\label{Sec:Sensitivity}
The measurement of the phase shift $\delta \phi \sim 10^{-10}$ (from \eqref{Eqn:PhaseShift}) is experimentally difficult. There are several factors that can however help measure such a phase shift. For example, even though the interrogation time $T$ of the experiment is at most the axion coherence time $\tau_{\text{max}} \sim 1 \text{ s } \l \frac{f_a}{\MGUT}\r$, the integration time of the experiment can be significantly longer. Even though the axion loses coherence after a time period  $\tau_{\text{max}}$, it will always remain spatially coherent over a distance $\sim \frac{1}{m_a \, v} \sim 600 \text{ km } \l \frac{f_a}{\MGUT}\r$. One can therefore construct two molecular interferometers, spaced within an axion wavelength $\sim \frac{1}{m_a \, v} \sim 600 \text{ km } \l \frac{f_a}{\MGUT}\r$ and cross-correlate their measurements over a long integration time. In this case, with an integration time $\tau_{\text{int}} \sim 10^6$ s, the required sensitivity of the molecular interferometer is 
\begin{equation}
\delta \phi_{\text{req}}  \sim \frac{10^{-7}}{\sqrt{\text{Hz}}} \, \l  \frac{T}{\text{1 s}}\r \, \l \frac{\delta E}{10^{-25} \text{ eV}} \r \l \sqrt{ \frac{\tau_{\text{int}}}{10^6 \, \text{s}}}\r
\label{Eqn:ShotNoise}
\end{equation}
The fact that the axion signal must be the same in two separated experiments differentiates it from backgrounds and allows this gain in sensitivity with integration time.

Another extremely interesting feature of this setup is the fact that the energy shift  \eqref{Eqn:EnergyShift} is independent of $f_a$. This feature arises from two facts: first, since the experiment measures a phase shift, it is sensitive to matrix elements which are suppressed by a single power of $f_a$. Second, the coupling between the axion and the gluon is a non-derivate coupling and hence the energy shift produced by the interaction is independent of the size of the apparatus. When combined with the fact that an axion with a higher $f_a$ has a larger coherence time $\tau_{\text{max}}$ (see equation \eqref{Eqn:MaxTime}), a significantly larger phase shift than the estimate \eqref{Eqn:PhaseShift} can be produced in the interferometer when searching for axions with $f_a \gg \MGUT$. This of course requires the ability to operate the interferometer for a time $T \gg 1$ s. 

Operation of the interferometer for times $T \gg 1$ s cannot be realized if the molecules are in free fall during the course of the experiment. However, if they are contained in a suitable trap with  trap lifetime significantly larger than 1 s, then the limit on the interrogation time would be set by this trap lifetime. For example, if the molecules can be trapped for $\sim 100$ s,  the sensitivity required to search for axions with $f_a$ between $\sim 10^{18} \text{ GeV } - \MPl$ would be 

\begin{equation}
\delta \phi_{\text{req}}  \sim \frac{10^{-5}}{\sqrt{\text{Hz}}} \, \l  \frac{T}{\text{100 s}}\r \, \l \frac{\delta E}{10^{-25} \text{ eV}} \r \l \sqrt{ \frac{\tau_{\text{int}}}{10^6 \, \text{s}}}\r
\label{Eqn:MoreTShotNoise}
\end{equation}

Thus, in an experiment where the molecules can be trapped for significantly longer than $\sim 1$ s, axions with high $f_a \sim \MPl$ are more visible than axions with lower $f_a$. This is an extremely interesting feature of this proposal and is in complete contrast  to other axion search methods where the signal  drops dramatically with increasing $f_a$.

\subsection{Backgrounds}
\label{Sec:Backgrounds}
The measurement of energy shifts as small as $\sim 10^{-25} \text{ eV} - 10^{-24} \text{ eV}$ (see equation \eqref{Eqn:EnergyShift}) requires technology that is sensitive enough to detect these tiny energy changes as well as control of backgrounds to the shot noise limit of the interferometer (see equation \eqref{Eqn:ShotNoise}). A complete analysis of all possible backgrounds and strategies to suppress them to the required level is beyond the scope of this paper. However, we note that the time varying nature of the axion induced phase shift can be successfully exploited to combat noise sources that typically arise in experiments that search for static, nuclear electric dipole moments. 

In the proposed detection strategy, the interfering states differ only by the relative orientation between the nuclear spin and the molecular axis. Both the states are placed in identical electronic configurations. Consequently, a variety of backgrounds that couple to the electronic states, for example, stray electric fields, will be common to both interfering states and will therefore not cause a phase shift. The most significant difficulties will arise from time varying backgrounds that couple directly to the nuclear spin. An example of such a background would be a time varying magnetic field. Naively, the power in the random, time varying magnetic field at the axion frequency $m_a \sim \text{ MHz } \left(\frac{10^{16} \text{ GeV}}{f_a}\right)$ has to be smaller than $\sim 10^{-18} \frac{\text{T}}{\sqrt{\text{Hz}}} \, \left(\sqrt{\frac{f_a}{10^{16} \text{ GeV}}}\right)$ to achieve the shot noise limit  \eqref{Eqn:ShotNoise}. 

The level of control required can be significantly ameliorated by operating the interferometer with another atomic or molecular species to serve as a comagnetometer. The signal induced by the axion is very sensitive to the choice of nucleus and molecular state and will affect the signal in the two species very differently. However, backgrounds like stray magnetic fields will be common and hence a measurement of the differential phase shift between the two species will significantly cancel such backgrounds, without affecting the signal. For example, if the interferometer is operated with a molecule containing Radium, the Radium atom can be used as a comagnetometer. In this case, the signal from the axion in the Radium atom is nearly zero while the Radium molecule will register a large effect (see Section \ref{Sec: overview}). However, the effect of a stray magnetic field on the Radium nucleus in both the molecule and the atom will be identical and hence will cancel in the differential measurement.  These species could then serve as a comagnetometer for the proposed experiment.  We note that atomic magnetometers with comparable sensitivities $\sim 10^{-16} \frac{\text{T}}{\sqrt{\text{Hz}}}$ have been demonstrated and it is believed that magnetometers as sensitive as $\sim 10^{-18} \frac{\text{T}}{\sqrt{\text{Hz}}}$ are achievable \cite{Romalis, RomalisPaper, RomalisPaper2}.

We also note that the axion oscillations are in the frequency range kHz - MHz, where these backgrounds may be naturally smaller than the static magnetic field background requirements that affect searches for static nucleon electric dipole moments.  The time varying nature of the axion signal also helps combat another major background that affects current molecular interferometry searches of static electric dipole moments. In these searches, the electric field $E$ used to polarize the molecule couples to the molecular velocity $V_{m}$ giving rise to a magnetic field $\sim E \, V_{m}$ in the rest frame of the molecule. This effective magnetic field couples to the nuclear magnetic moment and gives rise to energy and phase shifts. Combating this background requires significant cooling of the molecules. However, in this experiment, this static, effective magnetic field will not lead to significant phase shifts because its effects will average out during the induced nuclear spin precession necessary to see the time varying axion signal. 

We also  emphasize another key difference between this measurement strategy and resonant cavity axion searches. Both strategies require the tuning of experimental parameters to match the axion frequency. However, in a resonant cavity, this tuning is achieved by changing the physical size of the cavity, resulting in an enhancement in both the signal as well as other electromagnetic noise sources. With molecular interferometry, the tuning is attained by altering the frequency of the nuclear spin precession. This can be achieved by changing an external magnetic field, without amplifying other noise sources. 

In summary, while the search for the tiny energy shift caused by the dark matter axion requires careful consideration of backgrounds, the relatively high frequency nature of the axion signal as well as the intrinsic flexibility of atom and molecular interferometry ({\it e.g.}~the ability to use comagnetometers) should allow significant amelioration of background control requirements.  Further, the signal has a characteristic frequency given only by fundamental physics, and amplitude which scales in a known fashion with atomic species.  Thus, every different experiment, whether with the same or different atomic species, must observe the same frequency and even phase of the oscillating EDMs since these are properties of the axion dark matter.  This should allow significant background rejection and high confidence in a detection.

\subsection{Technology}
\label{Sec:Technology} 
Current molecular interferometers are not sensitive enough to reach the sensitivities  \eqref{Eqn:ShotNoise} and \eqref{Eqn:MoreTShotNoise}  required to search for the dark matter axion. The principal difficulty in extending the sensitivity of molecular interferometers has been the difficulty in cooling large molecular samples. Owing to their internal structure, traditional  cooling techniques ({\it e.g.} laser cooling) that are used to cool atoms cannot be directly applied to molecules. However, it is precisely these internal structures that make molecules incredibly valuable in the hunt for the dark matter axion. 

Considerable progress has been recently achieved in cooling molecular beams, thereby increasing the sensitivity of molecular interferometers. For example, the magnetic Feshbach resonance has been recently used to create $\sim 3 \times 10^5$ polar bialkali molecules of KRb \cite{UltracoldJunYe} cooled to sub microkelvin temperatures. These techniques could potentially be used to create bialkali molecules containing Francium, which belongs to the same chemical group as K and Rb. Innovative extensions of laser cooling techniques have also been successfully employed to create sub millikelvin ensembles of SrF \cite{DeMilleNature}, with possible applications  to other molecular structures. The molecular structure that allows for the successful laser cooling of SrF should also apply to other alkaline-earth monohydrides and monohalides \cite{DiRosa}. These techniques could prove particularly useful in dealing with molecules of Radium which is also an alkaline earth element. With these numbers, current molecular interferometers are between two to four orders of magnitude less sensitive than necessary (see equations \eqref{Eqn:ShotNoise} and \eqref{Eqn:MoreTShotNoise}). 

These sensitivities could improve considerably in the near future. The techniques employed by  \cite{UltracoldJunYe, DeMilleNature} are scalable and indeed, a variety of other molecular cooling techniques are being currently developed \cite{ColdMolecules}. Furthermore, it may also be possible to use squeezed molecular states to significantly enhance the shot noise limits of the interferometer. Such enhancements have been proposed for atomic interferometers \cite{squeezedstate} and an $\mathcal{O}\l 5\r$ enhancement has been recently demonstrated \cite{VladanHeisenberg}. 

In addition to developments in molecular interferometry, this proposal requires the use of actinide nuclei with octupole deformations which give rise to enhanced Schiff moments. These nuclei are radioactive and special techniques are needed to use them in the laboratory. Such techniques are currently being developed in order to use these nuclei to search for static nucleon electric dipole moments. In particular, successful laser trapping and cooling of Radium \cite{Scielzo:2006xc} and Francium \cite{Francium} atoms have been demonstrated. However, molecular states containing these actinides have not been used in laboratory studies. Further work is necessary to establish if the actinide nuclei produced by the above techniques can be successfully used to produce molecular states of interest. 

The detection of the dark matter axion field requires the harnessing of advances in both molecular interferometry and the ability to manipulate actinide nuclei. The current technological status in each of these areas is scalable and advances seem possible in the near future. Developments in each area is individually useful for a variety of physics searches, in particular the search for static nucleon electric dipole moments. However, the combination of these techniques will advance not only these dipole moment searches but also allow for the unique opportunity to search for the dark matter axion field.

\subsection{The Reach}
\label{Sec:Reach}
The energy shift \eqref{Eqn:EnergyShift} caused by the dark matter axion field is independent of $f_a$. The maximum phase difference \eqref{Eqn:PhaseShift} that can be caused by this energy shift depends entirely on the coherence time $T$ of the axion field. The larger this coherence time, the larger is the phase difference caused by this energy shift. This coherence time $T \sim \frac{1}{m_a \, v^2} \propto 1 \, \text{s} \, \l \frac{f_a}{10^{16} \text{ GeV}} \r$, leading to larger phase shifts for higher $f_a$. Furthermore, the axion oscillates at a frequency $m_a \sim \text{MHz } \l \frac{10^{16} \text{ GeV}}{f_a}\r$ which is faster by $\sim v^{-2} \sim 10^6$ than the maximal coherence time $T$.  In order to continuously accrue the phase difference over several axion oscillations, the relative orientation between the nuclear spin and the molecular axis must also oscillate at the axion frequency. This relative orientation can be more easily changed when the axion mass $m_a$  is smaller - thereby making it easier to search for higher $f_a$ axions. 

Owing to these reasons, this experimental strategy can be most easily implemented to search for axions in the range between the GUT ($\sim 10^{16}$ GeV) and the Planck ($\sim 10^{19}$ GeV) scales (see figure \ref{Fig:parameterspace}). The strategy can also potentially be extended to axions below the GUT scale, but this would require more sensitive molecular interferometers and the ability to more rapidly swap the orientation between the nuclear spin and the molecular axis.

\section{Conclusions and Future Vision}
\label{Sec:Conclusions}

We have proposed an experiment to detect cosmic axion dark matter.  This relies on the rapidly oscillating nuclear electric dipole moment that the axion generates.  This nuclear EDM couples with the internal electric fields in the atom, shifting the atomic energy levels.  These oscillating energy shifts can be searched for using molecular interferometry.  A discovery could be easily confirmed by using a different element because the signal has a characteristic frequency given only by fundamental physics, and amplitude which scales in a known fashion with atomic species.  Further, every different experiment, whether with the same or different atomic species, must observe the same frequency and even phase of the oscillating EDMs since these are properties of the axion dark matter.  This should allow significant background rejection and high confidence in a detection.

The molecular interferometry proposal discussed in this paper is most suitable for axion searches in the theoretically well motivated region of $f_a$ between the GUT ($\sim 10^{16}$ GeV) and Planck ($\sim 10^{19}$ GeV) scales.  Possibly it could be used to probe even lower scales down to maybe $\sim 10^{15}$ GeV. Current axion search strategies face severe limitations in probing this region. The use of molecular interferometry can potentially overcome several difficulties faced by current strategies - both in terms of the size of the signal with increasing $f_a$ and the ease with which the experiment can scan for the unknown axion mass.  While current  molecular interferometers fall short of the sensitivities needed to detect the dark matter axion field by two to three orders of magnitude, there are several avenues for improving this sensitivity which are already being actively explored for other reasons.

There are difficult technological challenges that must be overcome in order to transform this proposal into an experiment with the sensitivity necessary to detect the axion. This requires the development of techniques to cool and control molecules containing octupole deformed nuclei. Scalable technological advances appear to be possible in both the fields of molecular cooling and the manipulation of actinide nuclei which have octupole deformations. While developments in both these fields can individually be used to probe fundamental physics like the static nucleon electric dipole moment, the harnessing of these technologies opens up a unique opportunity to search for the cosmological axion field. 

The scalable nature of this molecular interferometry technique makes this proposal similar to the initial attempts to detect WIMP dark matter \cite{Goodman:1984dc}. These attempts were concerned with detecting WIMP nucleon cross-sections $\sim 10^{-38} \text{ cm}^2$. The scalable nature of the technology and the effort of a variety of experimental groups have now limited WIMP nucleon cross-section  $\lessapprox 10^{-44} \text{ cm}^2$ \cite{Ahmed:2009zw}, an improvement by over six orders of magnitude over the past two decades.  A similar experimental effort could transform this proposal into an experiment with the sensitivity to detect the cosmological axion field. Such an effort is  warranted in the case of the axion since it is a very well motivated dark matter candidate with very few feasible detection strategies, particularly in the region $f_a \gtrapprox 10^{16} \text{ GeV}$ (GUT) scale. 
 
The detection of the cosmological axion field would have an enormous impact on physics.  It would of course be the discovery of the nature of dark matter, a goal sought for many years.  In addition, it would allow a measurement of the local dark matter field, leading to the identification of its density and velocity components. In turn, these inputs may help us understand the evolution of dark matter in the Universe, particularly in the non-linear regime of galaxy formation.  Further, in addition to the identification of dark matter, it would herald the discovery of a new fundamental particle, the axion - leading to a resolution of the strong CP problem.  Perhaps most interestingly, the measurement of the axion's mass and couplings offers a rare glimpse into physics at very high energies  $ \gg 10^{12}$ GeV, far beyond the scales that can be probed directly in the laboratory.

\section*{Acknowledgments}

We would like to thank  Asimina Arvanitaki, Savas Dimopoulos, Sergei Dubovsky, Roni Harnik,  Bob Jaffe, David E. Kaplan, Mark Kasevich, Robert Laughlin, Peter Michelson,  Maxim Pospelov, John March-Russell,  John McGreevy, Xiao-Liang Qi and Frank Wilczek for useful discussions. S.R. would also like to thank the Center for Theoretical Physics, MIT for support under DOE Office of Nuclear Physics  grant DE-FG02-94ER40818 and NSF grant PHY-0600465 during the initial stages of this work.


\begin{thebibliography}{10}
\expandafter\ifx\csname url\endcsname\relax
  \def\url#1{{\tt #1}}\fi
\expandafter\ifx\csname urlprefix\endcsname\relax\def\urlprefix{URL }\fi

\bibitem{Clowe:2006eq}
  D.~Clowe, M.~Bradac, A.~H.~Gonzalez, M.~Markevitch, S.~W.~Randall, C.~Jones and D.~Zaritsky,
  Astrophys.\ J.\  {\bf 648}, L109 (2006)
  [arXiv:astro-ph/0608407].
  
\bibitem{Goodman:1984dc}
  M.~W.~Goodman, E.~Witten,
  Phys.\ Rev.\  {\bf D31}, 3059 (1985).
  
   
   \bibitem{clocks}
G.~Santarelli, {\em et~al.\/},
\newblock Phys. Rev. Lett. {\bf 82}, 4619 (1999).
J. McGuirk, et al., Opt. Lett. 26, 364 (2001).
S. Bize, et al., J. Phys. B: At. Mol. Opt. Phys. 38 (2005) S449-S468.

\bibitem{Metcalf}
H.~J.~Metcalf and P.~Straten, {\em Laser Cooling and Trapping\/} (New York: Springer, 1999).
  
\bibitem{squeezedstate}
M.~Kitagawa and M.~Ueda, Phys. Rev. A {\bf 47}, 5138 (1993).

\bibitem{Tuchman_PRA}
A.~K.~Tuchman {\em et al.}, Phys. Rev. A {\bf 74}, 053821 (2006).


\bibitem{Phillips2002:JPhysB}
J.~H. Denschlag, {\em et~al.\/},
\newblock J. Phys. B: At. Mol. Opt. Phys. {\bf 35}, 3095 (2002).

\bibitem{HolgerLMT}
H. Mueller, S.-w. Chiow, Q. Long, S. Herrmann, S. Chu., arXiv:0712.1990v1.

\bibitem{McGuirk}
J.M.~McGuirk,  {\em et~al.\/}, Phys. Rev. Lett. {\bf 85} 4498 (2000).

\bibitem{ColdMolecules}
L.~D.~Carr, D.~DeMille {\it et al.},
   [arXiv:0904.3175 [quant-ph]].

  
\bibitem{Peccei:1977hh}
  R.~D.~Peccei, H.~R.~Quinn,
  Phys.\ Rev.\ Lett.\  {\bf 38}, 1440-1443 (1977).
  
\bibitem{Peccei:1977ur}
  R.~D.~Peccei, H.~R.~Quinn,
  Phys.\ Rev.\  {\bf D16}, 1791-1797 (1977).
  
\bibitem{Weinberg:1977ma}
  S.~Weinberg,
  Phys.\ Rev.\ Lett.\  {\bf 40}, 223-226 (1978).
  
\bibitem{Wilczek:1977pj}
  F.~Wilczek,
  Phys.\ Rev.\ Lett.\  {\bf 40}, 279-282 (1978).
  
\bibitem{Kim:1979if}
  J.~E.~Kim,
  Phys.\ Rev.\ Lett.\  {\bf 43}, 103 (1979).
  
\bibitem{Shifman:1979if}
  M.~A.~Shifman, A.~I.~Vainshtein, V.~I.~Zakharov,
  Nucl.\ Phys.\  {\bf B166}, 493 (1980).
  
\bibitem{Dine:1981rt}
  M.~Dine, W.~Fischler, M.~Srednicki,
  Phys.\ Lett.\  {\bf B104}, 199 (1981).
  
\bibitem{Zhitnitsky:1980he}
  A.~R.~Zhitnitsky,
  Sov.\ J.\ Nucl.\ Phys.\  {\bf 31}, 529-534 (1980).
  
    
\bibitem{Pospelov:1999ha}
  M.~Pospelov, A.~Ritz,
  Phys.\ Rev.\ Lett.\  {\bf 83}, 2526-2529 (1999).
  [hep-ph/9904483].
  
 
  
\bibitem{Svrcek:2006yi}
  P.~Svrcek, E.~Witten,
  JHEP {\bf 0606}, 051 (2006).
  [hep-th/0605206].



  
  
\bibitem{Dicus:1978fp}
  D.~A.~Dicus, E.~W.~Kolb, V.~L.~Teplitz {\it et al.},
  Phys.\ Rev.\  {\bf D18}, 1829 (1978).
  
\bibitem{Vysotsky:1978dc}
  M.~I.~Vysotsky, Y.~.B.~Zeldovich, M.~Y.~.Khlopov {\it et al.},
  Pisma Zh.\ Eksp.\ Teor.\ Fiz.\  {\bf 27}, 533-536 (1978).
  
\bibitem{Raffelt:2006cw}
  G.~G.~Raffelt,
  Lect.\ Notes Phys.\  {\bf 741}, 51-71 (2008).
  [hep-ph/0611350].
  
\bibitem{Preskill:1982cy}
  J.~Preskill, M.~B.~Wise, F.~Wilczek,
  Phys.\ Lett.\  {\bf B120}, 127-132 (1983).
  
\bibitem{Abbott:1982af}
  L.~F.~Abbott, P.~Sikivie,
  Phys.\ Lett.\  {\bf B120}, 133-136 (1983).
  
\bibitem{Dine:1982ah}
  M.~Dine, W.~Fischler,
  Phys.\ Lett.\  {\bf B120}, 137-141 (1983).
  
\bibitem{Freivogel:2008qc}
  B.~Freivogel,
  JCAP {\bf 1003}, 021 (2010).
  [arXiv:0810.0703 [hep-th]].

  
\bibitem{Linde:1987bx}
  A.~D.~Linde,
  Phys.\ Lett.\  {\bf B201}, 437 (1988).
  
\bibitem{Sikivie:1985yu}
  P.~Sikivie,
  Phys.\ Rev.\  {\bf D32}, 2988 (1985).
  
\bibitem{Asztalos:2009yp}
  S.~J.~Asztalos {\it et al.}  [The ADMX Collaboration],
  Phys.\ Rev.\ Lett.\  {\bf 104}, 041301 (2010)
  [arXiv:0910.5914 [astro-ph.CO]].
  
\bibitem{Arvanitaki:2009fg}
  A.~Arvanitaki, S.~Dimopoulos, S.~Dubovsky {\it et al.},
   [arXiv:0905.4720 [hep-th]].
   
   
\bibitem{Banks:2003sx}
  T.~Banks, M.~Dine, P.~J.~Fox and E.~Gorbatov,
  JCAP {\bf 0306}, 001 (2003)
  [arXiv:hep-th/0303252].
   
   
   
\bibitem{'tHooft:1976up}
  G.~'t Hooft,
  Phys.\ Rev.\ Lett.\  {\bf 37}, 8-11 (1976).
  
\bibitem{'tHooft:1976fv}
  G.~'t Hooft,
  Phys.\ Rev.\  {\bf D14}, 3432-3450 (1976).

\bibitem{Hinds:1976ph}
  E.~A.~Hinds, C.~E.~Loving, P.~G.~H.~Sandars,
  Phys.\ Lett.\  {\bf B62}, 97-99 (1976).
  
\bibitem{Flambaum:1999zz}
  V.~V.~Flambaum,
  Phys.\ Rev.\  {\bf A60}, R2611-R2613 (1999).
  
\bibitem{Schiff}
  L.~I.~Schiff,
  Phys.\ Rev.\  {\bf 132}, 2194-2200 (1963).
  [hep-ph/9904483].
  
\bibitem{Flambaum:1984fb}
  V.~V.~Flambaum, I.~B.~Khriplovich, O.~P.~Sushkov,
  Sov.\ Phys.\ JETP {\bf 60}, 873 (1984).  
  
\bibitem{Dmitriev:2004gk}
  V.~F.~Dmitriev, R.~A.~Senkov,
  Phys.\ Atom.\ Nucl.\  {\bf 67}, 1799-1803 (2004).
  
\bibitem{Auerbach:1996zd}
  N.~Auerbach, V.~V.~Flambaum, V.~Spevak,
  Phys.\ Rev.\ Lett.\  {\bf 76}, 4316-4319 (1996).
  [nucl-th/9601046].
  
\bibitem{Spevak:1996tu}
  V.~Spevak, N.~Auerbach, V.~V.~Flambaum,
  Phys.\ Rev.\  {\bf C56}, 1357-1369 (1997).
  [nucl-th/9612044].

\bibitem{Scielzo:2006xc}
  N.~D.~Scielzo, I.~Ahmad, K.~Bailey {\it et al.},
  AIP Conf.\ Proc.\  {\bf 842}, 787-789 (2006).
  
\bibitem{Engel:1999np}
  J.~Engel, J.~L.~Friar, A.~C.~Hayes,
  Phys.\ Rev.\  {\bf C61}, 035502 (2000).
  [nucl-th/9910008].
  
  
\bibitem{Dimopoulos:2006nk}
  S.~Dimopoulos, P.~W.~Graham, J.~M.~Hogan {\it et al.},
  Phys.\ Rev.\ Lett.\  {\bf 98}, 111102 (2007).
  [gr-qc/0610047].
  
\bibitem{Dimopoulos:2007cj}
  S.~Dimopoulos, P.~W.~Graham, J.~M.~Hogan {\it et al.},
  Phys.\ Lett.\  {\bf B678}, 37-40 (2009).
  [arXiv:0712.1250 [gr-qc]].
  
\bibitem{Dimopoulos:2008hx}
  S.~Dimopoulos, P.~W.~Graham, J.~M.~Hogan {\it et al.},
  Phys.\ Rev.\  {\bf D78}, 042003 (2008).
  [arXiv:0802.4098 [hep-ph]].
  
\bibitem{Dimopoulos:2008sv}
  S.~Dimopoulos, P.~W.~Graham, J.~M.~Hogan {\it et al.},
  Phys.\ Rev.\  {\bf D78}, 122002 (2008).
  [arXiv:0806.2125 [gr-qc]].
  
  \bibitem{Romalis}
  http://physics.princeton.edu/romalis/magnetometer/
  
  \bibitem{RomalisPaper}
  J.~C.~Allred {\it et al.},
  Phys.\ Rev.\ Lett.\ {\bf 89}, 130801 (2002).

  \bibitem{RomalisPaper2}
  I.~M.~Savukov {\it et al.},
  Phys.\ Rev.\ Lett.\ {\bf 95}, 063004 (2002).
  
  \bibitem{UltracoldJunYe}
https://jilawww.colorado.edu/YeLabs/research/cold.html

\bibitem{DeMilleNature}
  E.~S.~Shuman, J.~F.~Barry, D.~DeMille,
  Nature, 467, 820-823 (2010)
  
\bibitem{DiRosa}
  M.~D.~Di~Rosa,
  Eur.\ Phys. \ J. \ D {\bf 31} 395-402 (2004)
  
\bibitem{VladanHeisenberg}
  I.~D.~Leroux, M.~H.~Schleier-Smith, V.~Vuletic,
  Phys.\ Rev. \ Lett. {\bf 104} 073602 (2010)
  
\bibitem{Francium}
  G.~D.~Sprouse, L.~A.~Orozco, J.~E.~Simsarian, W.~Shi, W.~Z.~Zhao
  Nucl.\ Instr.\ Meth.\ Phys.\ Res.\ B {\bf 126}, 370 (1997).  

\bibitem{Ahmed:2009zw}
  Z.~Ahmed {\it et al.} [ The CDMS-II Collaboration ],
  Science {\bf 327}, 1619-1621 (2010).
  [arXiv:0912.3592 [astro-ph.CO]].




\end{thebibliography}
\end{document}